\documentstyle[preprint,aps,prl,epsf]{revtex}

\input{epsf}
\input epsf.sty

\begin{document}
\title{STREAMER PROPAGATION IN MAGNETIC FIELD}
\author{V.N.Zhuravlev, T.Maniv}
\address{Department of Chemistry. Technion, Haifa 32000, ISRAEL}
\author{I.D.Vagner and P.Wyder}
\address{Grenoble High Magnetic Laboratory \\
Max-Planck-Institute f\"ur Festkorperforschung and\\
Center National de la Recherche Scientific,\\
\ \ 25 Avenue des Martyres, F-38042, Cedex 9, FRANCE \\
.}
\date{August 15, 1997}
\maketitle

\begin{abstract}
The propagation of a streamer near an insulating surface under the influence
of a transverse magnetic field is theoretically investigated. In the weak
magnetic field limit it is shown that the trajectory of the streamer has a
circular form with a radius that is much larger than the cyclotron radius of
an electron. The charge distribution within the streamer head is strongly
polarized by the Lorentz force exerted perpendicualr to the streamer
velocity. A critical magnetic field for the branching of a streamer is
estimated. Our results are in good agreement with available
experimental data.

\hspace{1.0in}

\end{abstract}

Recent experiments \cite{uhlig} on gas breakdown near an insulating surface
in a high magnetic field $\overrightarrow{B}$ have shown new remarkable
properties of such discharges. The channel of discharge in a magnetic field
appears to have a circular form with radius $R_s$ several orders of
magnitude larger than the electronic cyclotron radius. It decreases with
increasing $B$ and reaches $R_s$ $\sim 1cm$ at $B$ $\sim 7T$. At higher
magnetic field the discharge has a branched structure. These experiments
have shown that the streamer propagation cannot be treated as the motion of
a charge particle in crossed external electric ${\cal {E}}$ and magnetic $B$
fields.

Interest in the theory of streamers is usually associated with 
investigation of gas breakdown phenomena. On an
insulating surface near a point electrode, in the region with strong
electric field, the discharge has a filamentary structure. The tip of the
filament moves with high velocity $v_0$ ( $v_0$ $\sim 10^8cm/s$ ), that
exceeds the drift velocity $v_d$ of electrons in the streamer head field $%
\overrightarrow{E}_s$. The increase of external electric field $%
\overrightarrow{{\cal {E}}}$ results in penetration of some filaments ( so
called ''leaders'' ) deep into the surrounding gas. Although the plasma
parameters in streamers are different from those in leaders, their
propagation is associated with the same physical processes and is defined
mainly by the parameters of the streamer head. Since we are interested in
the behavior of the streamer front only, we will not distinguish here
between a streamer and a leader.

The streamer propagation mechanism was suggested by Raether, Loeb, and Meek
\cite{raether}-\cite{loeb}, and was further developed by other authors \cite
{turc}-\cite{tidman}. According to this theory the charged head induces in
its vicinity a strong electric field. This field leads to the increase of
the electron density ahead of the streamer front due to impact ionization .
The charge is displaced from this region via Maxwell relaxation. It is
assumed that the free electron density ahed of the streamer front is not 
zero due, e.g. to absorbtion of the streamer head radiation
produced, for example, by the streamer head radiation. The simple model
which takes into account only these main processes was considered by M.I.
D'yakonov and V.Y. Kashrovskii \cite{mid1}\cite{mid2}. They have estimated
theoretically the streamer parameters and have shown that the streamer
velocity $v_0$ and radius $r_s$  change smoothly with the external electric
field $\overrightarrow{{\cal {E}}}$, such that the propagation of the
streamer head can be treated as a quasistationary process.

In the present paper we generalize the streamer model \cite{mid1}\cite{mid2}
to include an external magnetic field. It is assumed that the plasma
filaments propagate in a plane perpendicular to the external magnetic field
and the streamer parameters do not change in the direction parallel to the
magnetic field. It is shown that , in the weak magnetic field limit, a
quasistationary streamer in the frame of reference rotating with a constant
angular velocity $\omega _s=v_0/R_s$ , proportional to the head charge
density, can be considered as a streamer in the absence of the magnetic
field . We estimate the main parameters of a streamer head, and show that
the obtained value for the radius of curvature is in close agreement with
experimental data \cite{uhlig}. The influence of the magnetic field on the
charge distribution within the streamer head is discussed, and a critical
magnetic field for the onset of branching is estimated and compared to the
experimental data.

Since the energy relaxation time of electrons is much larger than the
electron-ion relaxation time we ignore the gas heating processes. Thus the
concentration of atoms changes smoothly on the distance of the order of the
streamer head size and is assumed to be constant inside the head. We neglect
also the ion drift velocity in comparison with electron drift velocity $v_d$
and streamer velocity $v_0$.

The system of equations for the electron density $n$, the ion density $N$,
and the electric field $\overrightarrow{E}$ is
\[
\frac{\partial n}{\partial t}+div\left( n\overrightarrow{v}_d\right) =\beta
\left( E\right) n
\]
\[
\frac{\partial N}{\partial t}=\beta \left( E\right) n
\]
\begin{equation}
div\overrightarrow{E}=4\pi \rho \left( x\right) ,\text{ }rot\overrightarrow{E%
}=0  \label{e1}
\end{equation}
where $\rho \left( x\right) =e\left( N-n\right) $ is the charge
distribution, $\beta \left( E\right) =v_d\alpha \left( E\right) $. The
impact-ionization coefficient $\alpha \left( E\right) $ increases very
sharply with the field and saturates at some field value $E_0$ \cite{engel}
\begin{equation}
\alpha \left( E\right) =\alpha _0e^{-E_0/E}  \label{e2}
\end{equation}

We assume for simplicity that the electron drift velocity is proportional to
electric field $\overrightarrow{E}$, $v_{di}=\sum_{k}\mu _{ik}E_k$. 
Without external
magnetic field the mobility $\mu _{ik}$ is a diagonal tensor, $\mu _{ik}=\mu
_0\delta _{ik}$. In a weak magnetic field the mobility $\mu _{ik}$ is a
function of $\overrightarrow{B}$, which can be written as
\begin{equation}
\mu _{ik}=\frac{\mu _0}{1+\gamma ^2}\left( \delta _{ik}+\gamma \varepsilon
_{ik}\right) ,  \label{e3}
\end{equation}
where $\varepsilon _{ik}$ is the antisymmetric tensor in a plane
perpendicular to $\overrightarrow{B}$. The parameter $\gamma =\omega _B\tau
_{ea}$ , where $\omega _B=\frac{eB}{m_ec}$ is the cyclotron frequency and $%
\tau _{ea}$ is the time of electron-atom collisions, is assumed to be small,
$\gamma \ll 1$. In what follows we will be interested only in linear
corrections in $\gamma $ to the solution of Eq. (\ref{e1}). Since the
parameters $\alpha _0,$ $E_0$, and $v_d$ depend on $\gamma ^2$ \cite{brown}
they are magnetic field independent, in our approximation.

Let us consider the streamer propagation equation (\ref{e1}) in the frame of
reference with the origin at $\overrightarrow{r}_h=\frac{\int
\overrightarrow{r}\rho \left( \overrightarrow{r},t\right) d\overrightarrow{r}%
}{\int \rho \left( \overrightarrow{r},t\right) d\overrightarrow{r}}$ and
rotating with an angular velocity $\omega \left( t\right) $ . The resulting
new coordinates are:
\begin{equation}
\eta _i=\sum_k\Omega _{ik}\left( t\right) \left( x_k-r_{hk}\left( t\right)
\right)   \label{e4}
\end{equation}
where $\Omega _{ik}\left( t\right) $ is the rotation matrix for the angle $%
\varphi =\int \omega \left( t\right) dt$ in the plane perpendicular to $%
\overrightarrow{B}$:
\begin{equation}
\Omega _{ik}\left( t\right) =\left(
\begin{array}{cc}
\cos \varphi  & \sin \varphi  \\
-\sin \varphi  & \cos \varphi
\end{array}
\right)   \label{e4a}
\end{equation}

The transformation of derivatives are
\begin{equation}
\frac \partial {\partial x_i}\rightarrow \sum_k\frac \partial {\partial \eta
_k}\Omega _{ki},\,\,\,\,\,\,\,\,\,\,\,\,\,\,\,\,\,\,\,\,\,\frac \partial {%
\partial t}\rightarrow \frac \partial {\partial t}+\sum_{ikl}\frac \partial {%
\partial \eta _k}\left[ \stackrel{.}{\Omega }_{ik}\Omega _{kl}^{-1}\eta
_l-\Omega _{ki}\stackrel{.}{r}_{hi}\right]   \label{e5}
\end{equation}
Here the dot denotes time derivative. It follows from Eq. (\ref{e5}) that
the streamer charge propagation is quasistationary if
$\sum\limits_k\stackrel{.}{\Omega }_{ik}\Omega
_{kl}^{-1}=\varepsilon _{il}\omega =const$ and $\sum\limits_i\Omega _{ki}%
\stackrel{.}{r}_{hi}=v_{0k}=const$. Thus the head of the quasistationary
propagating streamer moves with constant velocity $v_0$ along a circle with
radius $R_s=v_0/\omega $.

Rewriting Eq.(\ref{e1}) for quasistationary propagation in the rotating
frame we obtain
\[
\sum_{il}\frac \partial {\partial \eta _i}n\left[ -v_{0i}+\omega \varepsilon
_{il}\eta _l+\mu _0\left( E_i+\gamma \varepsilon _{il}E_{0l}\right) \right]
=\beta \left( E\right) n
\]
\[
-\sum_iv_{0i}\frac{\partial N}{\partial \eta _i}=\beta \left( E\right) n
\]
\begin{equation}
\sum_i\frac{\partial E_i}{\partial \eta _i}=4\pi \rho \left( \eta \right) ,%
\text{ }\sum_{ik}\varepsilon _{ik}\frac{\partial E_i}{\partial \eta _k}=0
\label{e6}
\end{equation}
Eq. (\ref{e6}) should be solved with the following boundary condition
\begin{equation}
\rho _sv_0=en_s\mu _0E_{s\text{,}}  \label{e7}
\end{equation}
which follows from the charge conservation on the surface of the streamer
front. Here $n_s$ and $\rho _s=e\left( N-n_s\right) $ are the electron and
the charge densities on the front.

Appearing in the first order term with $\gamma ,$ the electric field $E_{0k}$
is the field of a streamer propagating in the absence of external magnetic
field. It consists of two parts: the external field $\overrightarrow{{\cal {E%
}}}$ and field $\overrightarrow{E}_\rho $ created by the head charge. Field $%
\overrightarrow{E}_\rho $ is a symmetrical function with respect to the
streamer axis. Usually $\left| \overrightarrow{{\cal {E}}}\right| $ is
negligible in comparison with $\left| \overrightarrow{E}_\rho \right| $, but
near the electrode it can strongly influence the streamer propagation.

Let us expand $E_{0k}\left( \eta \right) $ near the central point $\eta _i=0$%
\begin{equation}
E_{0k}\left( \eta \right) =E_{0k}\left( 0\right) +a\sum_{l}\delta_{kl}
\eta_l+\sum_{l}b_{kl}\eta_l +D_k  \label{e8}
\end{equation}
where $a=2\pi \rho \left( 0\right) $, $b_{kl}$ - symmetrical matrix with $%
Sp\left( b\right) =0$. The term $E_{0k}\left( 0\right)+\sum_{l}b_{kl}
\eta_{l} $ corresponds to the potential field, which satisfies the
Laplace equation and can be absorbed into $E_k$ as a correction.
The field $D_k$, defined by Eq.(\ref{e8}), is
proportional to the deviation of the charge distribution from the uniform
one. It is small in the central region and becomes large near the surface of
the streamer head. Assuming that the streamer propagation is determined mainly
by the central region, one can discard the terms with
$\overrightarrow{D}$ while determining the streamer trajectory.

The second term in (\ref{e8}) leads to the streamer curving. At $\omega
=-2\pi \rho \left( 0\right) \gamma \mu _0$ the equations (\ref{e6}) turn to
a system of equations describing the quasistationary streamer propagation
without magnetic field. Thus streamers moving from cathode or anode will
curve in opposite directions with frequency $\omega _s=\left| \omega \right|
=$ $\left| 2\pi \rho \left( 0\right) \gamma \mu _0\right| $.

Introducing Maxwell relaxation time $\tau _m^{-1}=4\pi \mu _0en_s$, one
obtains $\omega _s\tau _m=\frac{\gamma \rho \left( 0\right) }{2en_s}$. If
the streamer radius $r_s$ is of the same order of magnitude as the
characteristic distance of the increase of electric field from internal
region to the front, one can estimate $r_s$ as $r_s\simeq \tau _mv_0$. So we
have
\begin{equation}
\frac{r_s}{R_s}=\frac{\gamma \rho \left( 0\right) }{2en_s}  \label{e9}
\end{equation}
This value is very small, since $\gamma \ll 1$ and $\frac{\rho \left(
0\right) }{en_s}\sim \frac{\mu E_s}{v_0}\ll 1$. The last inequality can be
obtained from Eq. (\ref{e7}) at $\rho \left( 0\right) \simeq \rho _s$.
Parameter $\gamma $ is proportional to magnetic field $B$, so the streamer
radius decreases as $1/B$. This form of $R_s\left( B\right) $ is somewhat
different from the experimentally observed field dependence reported in
\cite{uhlig}$, i.e. R_s\left( B\right)\sim 1/B^\alpha $,
where $\alpha \sim 1.3-1.5$. Such a disagreement is
connected probably with the approximate description of the ionization
coefficient and mobility by formulas (\ref{e2}), (\ref{e3}). It must be
especially noticeable at high magnetic field.

To evaluate the streamer head charge we will consider the one dimensional
streamer equations (\ref{e1}). Such approximation holds if the width $\delta
$ of the streamer front is much smaller than the head size $r_s$: $\delta
\ll r_s$. This is the case if the electron density in front of the streamer
is much smaller than inside \cite{mid2}. Equations (\ref{e1}) at $B=0$ have
a simple analytical solution. Assuming that the streamer moves along the
$x$-axis and choosing the boundary conditions as $n\left( -\infty \right)
=n_\infty $, \ $E\left( -\infty \right) =0,$ and $n\left( E=E_s\right) =0$
we can easily obtain the relation between the equilibrium electron density $%
n_\infty $ and the electric field $E_s$ on the front
\begin{equation}
\frac{n_\infty }{n_0}=\frac 1{1-\frac{\mu _0E_s}{v_0}}\int%
\limits_0^{E_s/E_0}e^{-1/x}dx  \label{e10}
\end{equation}
Here $n_0=\frac{\alpha _0}{4\pi e}$. This solution describes a plane wave
with narrow front if $\mu _0E_s\ll v_0$. In the opposite case $\mu
_0E>>v_0$,one can neglect time derivatives in (\ref{e1}) and obtain
the stationary solution.

The equilibrium electron density $n_\infty $ and the propagation velocity $%
v_0$ are defined by the conditions of the streamer formation. This stage of the
discharge development should be described by essentially nonstationary
equations. Their solution depends on external electric field and parameters
of initial ''seed''. On the quasistationary stage of the streamer
propagation the electron density $n_\infty $ can be estimated with the help
of the relation
\begin{equation}
r_s\simeq \frac{v_0}{4\pi e\mu _0n_s}  \label{e11}
\end{equation}
Here it is supposed that $n_\infty \simeq n_s$ in agreement with $\rho
\left( 0\right) \sim \rho _s\ll en_s$. Equation (\ref{e7}) allows to relate
the charge density $\rho _s$ with $E_s$. Note that $E_s$ and $\rho _s/en_s$
have logarithmic dependence on the density $n_s$ and radius $r_s$. Thus, the
experimental error in $r_s$ gives rise to small logarithmic correction
to the relative charge $\rho _s/en_s$.

Let us now compare our results with the experiment. For the streamer and
plasma parameters in the absence of a magnetic field we have used the data
from the paper of Dhali and Williams \cite{dhali} for the streamer in $N_2$
at atmosphere pressure: $v_0=2\cdot 10^8cm/s$, $r_s\simeq 10^{-2}cm$.
Substituting these values to (\ref{e10}) and (\ref{e11}) one obtains $%
n_s\simeq 3\cdot 10^{13}cm^{-3}$, $E_s/E_0\simeq 0.6$, $\rho _s/en_s\simeq
\mu _0E_s/v_0\simeq 0.2$. Estimating $\gamma =0.04$ at $B=1T$ we have from
Eq. (\ref{e9}) for the trajectory curvature radius $R_s\simeq 0.5cm$ at $%
B=5T $. The experimental value of $R_s$ \cite{uhlig} for the same conditions is
slightly larger, i.e. $R_s^{ex}\simeq 1.2cm$. This discrepancy can be explained
e.g. by the growth of the charge density from external region towards the
streamer front.

Let us consider the streamer in the limit of an infinitely narrow front in
the system of reference where the streamer head is at rest . The electric
field $E$ inside the head is sufficiently small so that the ionization
process can be safely neglected. The ions are assumed not to be affected by
the electromagnetic field, i.e. $N=const$. The corresponding equations are
\[
\sum_{kl}\frac \partial {\partial \eta _k}\left[ n\left( -v_{0k}+\mu
_{kl}E_l+\omega \varepsilon _{kl}\eta _l\right) \right] =0
\]
\begin{equation}
\sum_k\frac{\partial E_k}{\partial \eta _k}=4\pi \left( N-n\right)
\label{e12}
\end{equation}
For simplicity the streamer body will be represented as a cylinder with
radius $r_s$ and axis directed along the $\eta _x$-axis. Writing the content
of the square brackets in Eq.(\ref{e12}) by:
\begin{equation}
\sum_ln\left( -v_{0k}+\mu _{kl}E_l+\omega \varepsilon _{kl}\eta _l\right)
=v_0N\frac{\partial \Phi }{\partial \eta _k}  \label{e13}
\end{equation}
the function $\Phi $ satisfies the Laplace equation $\Delta \Phi =0$ inside
the streamer body with boundary conditions
\begin{equation}
\frac{\partial \Phi }{\partial \eta _x}=1,\frac{\partial \Phi }{\partial
\eta _y}=0\text{ at }\eta _x=0\text{; }\frac{\partial \Phi }{\partial \eta _k%
}=0\text{ at }\eta _y=r_s\text{; }\Phi \rightarrow 0\text{ at }\eta
_x\rightarrow \infty  \label{e14}
\end{equation}

Using $E_l$ from (\ref{e13}) and substituting it into Poisson equation we
obtain the following equation for the normalized electron density $\overline{%
n}\equiv n/N$%
\begin{equation}
\left( \nu _{ik}\frac{\partial \Phi }{\partial \eta _k}\right) \frac{%
\partial \overline{n}}{\partial \eta _i}=-\frac{\overline{n}^2\left(
\overline{n}-1\right) }{L_0}+\frac{2\gamma \overline{n}}{R_s}  \label{e15}
\end{equation}
with the boundary condition
\begin{equation}
\overline{n}\left( \eta _x=0,\eta _y=0\right) =1+\Delta \overline{n}
\label{e16}
\end{equation}
where $\nu _{ik}=\delta _{ik}-\gamma \varepsilon _{ik}$, $L_0=\frac{v_0}{%
4\pi e\mu _0N}$ is the characteristic length of the charge relaxation and $%
\Delta \overline{n}=\rho _s/eN$. Since $\gamma \ll 1$ and $L_0\simeq r_s\ll
R_s$ the second term on the RHS of (\ref{e15}) may be neglected.
In the case of small
charge density $e\Delta \overline{n}\ll e\overline{n}$ the equation (\ref
{e15}) has a simple analytical solution
\begin{equation}
\overline{n}\left( \eta _x,\eta _y\right) =1+\Delta \overline{n}\exp \left( -%
\frac{s\left( \eta _x,\eta _y\right) }{L_0}\right)  \label{e17}
\end{equation}
where the effective path $s\left( \eta _x,\eta _y\right) $ is defined by
an integral over a path from the point $\left( 0,0\right) $ to the point $\eta
\equiv \left( \eta _x,\eta _y\right) $, i.e:
\begin{equation}
s\left( \eta _x,\eta _y\right) =\int\limits_{\left( 0,0\right) }^{\left(
\eta _x,\eta _y\right) }\frac{\sum\limits_{ik}\nu _{ik}\frac{\partial \Phi }{%
\partial \eta _k}d\eta _i}{\sum\limits_i\left| \sum\limits_k\nu _{ik}\frac{%
\partial \Phi }{\partial \eta _k}\right| ^2}  \label{e18}
\end{equation}

According to (\ref{e14}) near the front surface $\frac{\partial \Phi }{%
\partial \eta _i}=\delta _{ix}$, so that to the first order with $\gamma $,
\ $s\left( \eta \right) =\eta _x-\gamma \eta _y$ and
\begin{equation}
\overline{n}\left( \eta _x,\eta _y\right) =1+\Delta \overline{n}\exp \left( -%
\frac{\eta _x}{L_0}+\gamma \frac{\eta _y}{L_0}\right)  \label{e19}
\end{equation}
The corresponding electric field is
\[
E_{\eta _x}\left( \eta _x,\eta _y\right) =E_s\exp \left( -\frac{\eta _x}{L_0}%
+\gamma \frac{\eta _y}{L_0}\right)
\]
\begin{equation}
E_{\eta y}\left( \eta _x,\eta _y\right) =\gamma E_s\exp \left( -\frac{\eta _x%
}{L_0}+\gamma \frac{\eta _y}{L_0}\right)  \label{e20}
\end{equation}
where $E_s=e\Delta \overline{n}\frac{v_0}{\mu _0}$.

Thus, an electric field $E$ of the order of $E_s$,in the 
$\eta _x$-direction at the streamer front stimulates its propagation in
this direction. It is therefore reasonable to suggest that a
sufficiently strong electric field component $E_{\eta _y}$ $\sim E_s$,
perpendicular the streamer propagation, will lead to the breakdown
of the streamer head and the formation of a new streamer deflected along
the $\eta _y$-direction with respect to the original one. The new streamers
will arise only from one side in the plane transverse to $%
\overrightarrow{B}$ . Substituting $\eta _x=0$, $\eta _y=r_s\simeq L_0$ in (%
\ref{e20}) we conclude that the condition $E_{\eta _y}$ $\sim E_s$ is
fulfilled at $\gamma e^\gamma \sim 1$, i.e. $\gamma \simeq 0.6$. This value
for the atmosphere discharge in $N_2$ corresponds to $B=12T$, which
closely agree with the experimental result $B^{ex}\simeq 7T$ \cite{uhlig}.

In conclusion we have shown that the simple model taking into account only
the main processes provides a reasonably good description of the streamer
discharge in a magnetic field. The streamer head propagation is very similar
to the movement of a free charged ''particle''. Without magnetic field this
''particle'' moves with a constant velocity $v_0=const$. In the presence of
magnetic field the trajectory has a circular form. Such a simple picture
occurs when the external electric field ${\cal {E}}$ is negligible in
comparison with the field $E_s$ of the charged streamer head. Nevertheless,
the role of the external field ${\cal {E}}$ is very important not only for
maintenance of the discharge, but also for the definition of the streamer
parameters on the initial stage of the development. Strong electric field 
$\cal {E}\sim E_{s}\sim E_0$ distorts the circular trajectory making it
similar to the trajectory of a charged particle in crossed electric and
magnetic fields. This phenomenon was observed in \cite{uhlig}.

To estimate the ''particle'' mass density $\rho _m$ we compare the
expression for the radius $R_s$ in the form $R_s=\frac{\rho _mv_0c}{\rho _sB}
$ with Eq.(\ref{e9}). We obtain the following expression for the ratio of
the mass and the charge densities
\begin{equation}
\frac{\rho _m}{\rho _s}=\frac 2{\Delta \overline{n}}\frac{\tau _m}{\tau _{ea}%
}\frac{m_e}e  \label{e21}
\end{equation}
Because of the large parameter $\frac 2{\Delta \overline{n}}\frac{\tau _m}{%
\tau _{ea}}>>1$ the streamer head turns in magnetic field more slowly than
the particle with charge density $\rho _s$ and mass density $\rho _m\simeq
m_e\rho _s/e$, whose radius does not depend on the plasma parameters of the
streamer head.

ACKNOWLEDGMENT: We are grateful to P.Uhlig for useful discussions.
This research was supported by a grant from the US-Israeli Binational
Science Foundation grant no. 94-00243, by the fund for the promotion
of research at the Technion, and
by the center for Absorption in Science, Ministry of Immigrant Absorption
State of Israel.

\end{document}